\documentclass[12pt]{article}
\usepackage{graphics}
\usepackage{amsmath}
\usepackage{float}
\usepackage{epsfig}
\usepackage{epstopdf}
\usepackage{setspace}
\title{Bulk viscous Zel'dovich fluid model and it's asymptotic behavior}

\author{Rajagopalan Nair K and Titus K Mathew \\
Department of Physics, \\ Cochin University of Science and Technology, \\ Kochi-682022,
India \\
E-mail:rajagopal\_physics\_61@cusat.ac.in; titus@cusat.ac.in}
\date{}
\begin{document}

\maketitle

\abstract{
In this paper we have considered a flat FLRW universe with bulk viscous Zel'dovich as the cosmic component. Being considered the bulk viscosity
as per the Eckart formalism, we have analyzed the evolution of the Hubble parameter and constrained the model with the Type Ia Supernovae
data thus extracting the constant bulk viscous parameter and present Hubble parameter. Further we have analyzed the scale factor, equation of state and deceleration 
parameter. The model predicts the late time acceleration and is also compatible with the age of the universe as given by the oldest globular 
clusters. We have also studied the phase-space behavior of the model and found that a universe dominated by bulk viscous Zel'dovich fluid is stable. But on the 
inclusion of radiation component in addition to the Zel'dovich fluid, makes the model unstable. Hence, even though the bulk viscous Zel'dovich fluid 
dominated universe is a feasible one, the model as such failed to predict a prior radiation dominated phase.} 



%
%
\section{Introduction}
\label{intro}
Observational data on Type-Ia supernovae
\cite{Perl1,Reiss1,Hicken1,Shariff1,Koivisto1,Daniel1,Fedeli1} and
the CMB \cite{Komatsu1,Larson1} has confirmed with sufficient
accuracy that nearly seventy percent of energy of the universe is an
exotic form called dark energy which is responsible for the current
acceleration of the universe. The remaining part of the cosmic
components consists of nearly twenty three to twenty four percentage
of weakly interacting dark matter
\cite{Refregier1,Tyson1,Allen1,Zwicky1,Rubin1} and a few percentage
of luminous matter and radiation. Even with the overwhelming
evidence for the existence of these cosmic components the current
observational data does not rule out the possible existence of other
exotic fluid components. One of the example for such fluids is the
dark radiation, which can exist in the early or later stage during
the evolution of the universe \cite{Dutta1}. Recently, considerable
attention has been paid to the study of another exotic fluid, the
Zel'dovich fluid or stiff fluid, first studied by Zel'dovich
\cite{Zel'dovich1}. Zel'dovich fluid is a perfect fluid in which the
speed of sound is equal to the speed of light so that the equation
of state becomes $\omega_z = p_z/\rho_z =1,$ the highest value a
fluid can have in consistency with causality.

Zel'dovich fluid or stiff fluid behavior particularly in the cosmological context had
been considered by many. In dealing with the self interaction
between dark matter components, authors in reference \cite{Steile1}
have shown that the self interaction field will behave like a stiff
fluid. The existence of Zel'dovich fluid was confirmed in the
Horava-Lifshitz gravity based cosmological models, when the so
called detailed balancing conditions \cite{Horava1,Kiritsis1} were
relaxed \cite{Sotiriou1,Bogdanos1}. The relevance of the existence
of the Zel'dovich fluid in the early universe was discussed in
reference \cite{Barrow1}. In certain inhomogeneous cosmological
models stiff fluid was arised as an exact non singular solution
\cite{Fernandes1,Fernandes2}. In the standard evolution of the
Friedmann universe the density of the Zel'dovich fluid is found to
be decreasing faster than radiation and matter. So its effect on
early universe would be larger. One of the phenomena that took place
in the early universe is the primordial nucleosynthesis which might
be influenced by the presence of stiff fluid. In reference
\cite{Dutta2} the authors have found a limit on the density of the
stiff fluid from the constraints  on the abundances of the light
elements. Besides, there are no empirical facts in rebuttal to the
stiff fluid.

 In an expanding universe there arise
deviations from local thermodynamic equilibrium.  Consequently there
can arise bulk viscosity in the cosmic fluid which will restore the
equilibrium \cite{Wilson1}. This bulk viscosity modifies the
effective pressure of the fluid in order to facilitate regaining of
the equilibrium situation. As soon as the equilibrium is reached the
bulk viscous pressure vanishes \cite{Wilson1,Ilg1}. In the context
of inflation in the early universe it was already shown that an
imperfect fluid with bulk viscosity can cause the early accelerated
expansion \cite{Zimdahl1}. The late time viscous universe was
studied in reference \cite{Paddy1}. A considerable number of
studies by including bulk viscosity in the dark matter setter were
carried out by many in the context of the late acceleration of the
universe \cite{Avelino1,Avelino2,Athira1}.

Recently considerable interest have been shown in the study of
Zel'dovich fluid in an expanding universe. In reference
\cite{Titus1} the authors studied the evolution of viscous
Zel'dovich fluid in a flat universe and found that it can have
considerable effect even in the late universe. However, they have
n't tried to constrain the model with cosmological model
observational data to arrive at a realistic picture. In the present
work we are trying to compare the evolution of a flat universe with
bulk viscous Zel'dovich fluid with the latest cosmological data on
Type I a supernovae. We have evaluated the model parameters
including the transport coefficient of bulk viscosity and studied
the evolution particularly in the late stage of the universe. The
paper is organized as follows. In section 2, we derive the Hubble parameter and constrained it using 
type Ia Supernovae data to extract the constant bulk viscous parameter and the present value of the Hubble parameter. We also 
include the evolution of the equation of state parameter and deceleration in this section. In section 3, we present our analysis on 
the space-space structure of the model, followed by conclusions in section 4.

\section{The bulk viscous Zel'dovich fluid model}

The main feature of the Zel'dovich fluid is that sound velocity in
the fluid is equal to that of light. The equation of state
\cite{Zel'dovich1} is given by,
\begin{equation}\label{eqn:EquationofState}
p_z=\rho_z.
\end{equation}
A similar equation of state was studied with reference to some
special case by Masso and others \cite{Masso1}. In including the
viscosity in the fluid we will follow the Eckart formulation which
deals with the viscous dissipative processes occurred in a
thermodynamical system when it deviates from local equilibrium. An
equivalent formulation was developed by Landau and Lifshitz
\cite{Landau1}. However it was noted that the equilibria in Eckart's
frame are unstable \cite{Hiscock1} and signals were propagated
through the fluid at superluminal velocities \cite{Israel1}. These
draw backs were rectified in a more general formalism by Israel et
al.\cite{Israel2,Israel3} from which Eckart's theory follows as a
first order limit. But many authors are still using Eckart's theory
because of its simplicity. For example, Eckart's formalism was used
in some models on the late acceleration of the universe caused by
the bulk viscous dark matter \cite{Kremer1,Hu1,Ren1}. In the mean
time Hiscock et al. \cite{Hiscock2} have shown that Eckart formalism
can be favored over the Israel-Stewart formalism in inflation
during the early universe using bulk viscosity. In the present work
we too follow Eckart's approach so that the effective pressure of
the bulk viscous Zel'dovich fluid can be expressed as
\begin{equation}\label{eqn:effectivepressure}
p_z^{'}=p_z - 3\zeta H
\end{equation}
where $\zeta$ is the coefficient of viscosity and $H$ is the Hubble
parameter.

We consider a flat Freedmann universe with FLRW metric given as
\begin{equation}
ds^2=dt^2 - a^2(t)(dr^2 + r^2 d\theta^2+r^2 \sin^2\theta d\phi^2)
\end{equation}
where $t$ is the cosmic time, $a(t)$ is the scale factor of
expansion, $r,\theta,\phi$ are the comoving coordinates. This when
combined with the Einstein's field equations gives the dynamical
equations
\begin{equation}\label{eqn:EinsteinFlEq}
 H^2=\rho/3
\end{equation}
and the conservation equation
\begin{equation}\label{eqn:Conservation}
\dot{\rho} +3H(\rho +p)=0
\end{equation}
Here we follow the standard convention $8 \pi G=1$. When these
equations are combined the equation (\ref{eqn:effectivepressure})
for effective pressure gives,
\begin{equation}\label{eqn:Hevolution}
dH/dt-3H^2-\frac{3}{2} \zeta H=0
\end{equation}
Solving these equations after changing the variable from $t$ to
$x=\log a$ we get
\begin{equation}\label{eqn:Hubblepara1}
h=\frac{1}{6} \left(\bar\zeta+(6-\bar\zeta)a^{-3} \right),
\end{equation}
where $h=H/H_0, \,\, H_0$ is the present Hubble parameter and
$\bar\zeta=\zeta/H_0$ is the dimensionless viscous parameter. For
$\bar\zeta=0$ the Hubble parameter becomes $H \sim H_0 a^{-3}$. The
density of the Zel'dovich fluid will then evolve as $\rho \sim
a^{-6}$  and the scale factor will evolve as $a\sim (H_0 t)^{1/3}$
and hence the universe would have eternally decelerated and hence
the effect of the Zel'dovich fluid will be relevant to the early
epoch of the universe \cite{Dutta2}. On the other hand, the presence
of viscosity will directly contribute a negative term, the effect of
which will depend upon the value assumed by the viscous parameter
$\bar\zeta.$ If $\bar\zeta>6$ the scale factor always grows
exponentially with time, in other words, an eternal acceleration.
While for $0< \bar\zeta <6$ the scale factor shows an initial
deceleration followed by an acceleration in expansion in the later
phase. So the admissible values of $\bar\zeta$ is very important in
this model which is to be evaluated by the observational
constraints.

\subsection{Extraction of the model parameters using Type Ia
supernovae data}

The model parameter $\bar\zeta$ and the Hubble parameter $H_{0}$ can
be extracted using Type Ia supernovae data. We have used Union data
which consists of 307 data points \cite{Kovalsky1} in the redshift
range $ 0.01<z<1.6 $. The distance modulus of the supernova at a red
shift z is
\begin{equation}
\mu_i(z) =(m-M) = 5\log_{10} d_L(z)+25
\end{equation}
where $m$ is the apparent magnitude, $M$ is the absolute magnitude
and $d_L(z)$ is the luminosity distance of a supernova in a flat
FLRW universe, which is calculated with the relation,
\begin{equation}
d_L(z)=\frac{c(1+z)}{H_0} \int_0^z \frac{dz^{'}}{h(z^{'})}
\end{equation}
where $h(z^{'})$ is identical with the normalized Hubble parameter
given in equation(\ref{eqn:Hubblepara1}). The distance moduli of
\begin{figure}
\centering
\includegraphics[scale=0.75]{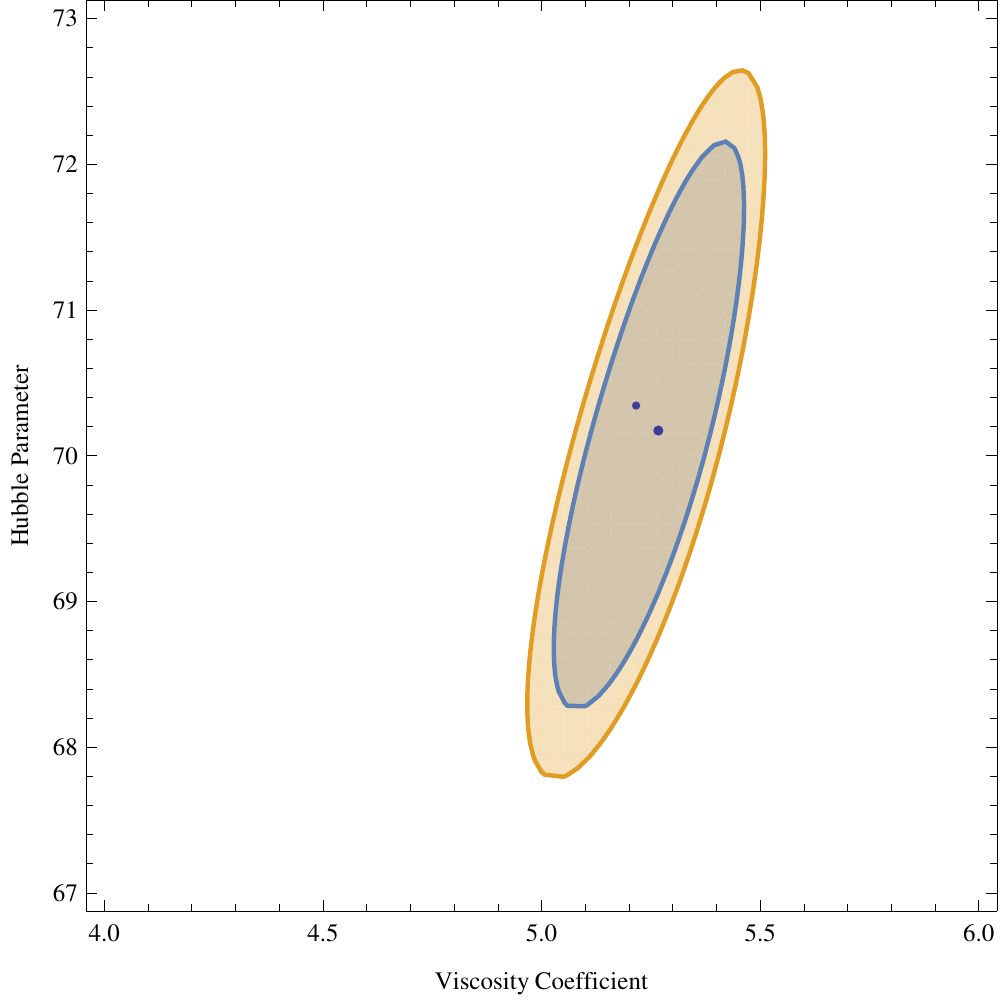}
\caption{Confidence intervals for the parameters $\bar\zeta$ and
$H_0.$ The outer curve corresponds to 99.99$\%$ probability and the
inner one corresponds to 99.73 $\%$ probability. The lower dot
represents the values of the parameters corresponding to the minimum
of the $\chi^2.$} \label{fig:contour}
\end{figure}
 supernovae at various red shifts are calculated and are
compared with the corresponding observational data. We then
construct the statistical $\chi^2$ function
\begin{equation}
\chi^2=\sum_{k=1}^n {\left[\mu_i-\mu_k \right]^2 \over \sigma_k^2}
\end{equation}
where $\mu_k$ is the observed distance modulus of the $k^{th}$
supernova, and $\sigma_k$ is the variance of the measurement and n
is the number of data points. The parameter values are obtained by
minimizing the $\chi^2$ function. The confidence regions in figure
\ref{fig:contour} for the parameters $\bar \zeta$ and $H_0$ are then
constructed for 99.73$\%$ and 99.99$\%$ respectively to find the
best estimate of the parameters. The values of the parameters are
shown in table \ref{table:T1}
\begin{table}[h]
\centering
\begin{tabular}{|c|c|c|c|c|}
\hline
Model & $\chi^2_{min}$ & $\chi^2_{min}/d.o.f. $ & $\bar\zeta$
& $H_0$
\\ \hline
Bulk viscous model & 300.264 & 1.011 & 5.25 & 70.20\\ \hline
$\Lambda$CDM model & 300.93 & 1.013& - & 70.03\\ \hline
\end{tabular}
\caption{The best estimates of the parameters $\bar\zeta$ and $H_0$
evaluated with the supernova type-Ia union data 307 data points. But
we avoided some low red shifts data, so that the net number of data
used is 297.} \label{table:T1}
\end{table}
and for a comparison we have also evaluated the parameters of
$\Lambda$CDM model. With the statistical correction the values of
parameters finally become, $\bar\zeta=5.25\pm 0.14$  and
$H_0=70.20\pm 0.58.$

\subsection{Evolution of cosmic parameters}

The behavior of scale factor in the Zel'dovich fluid dominated
universe can be obtained from the equation(\ref{eqn:Hubblepara1}) as
\begin{equation}
a(t)= \left((\bar\zeta -6)+6e^{\frac{H_0\bar\zeta}{2}(t-t_0)}
\over\bar\zeta\right)^{1/3}.
\end{equation}
At sufficiently early time the scale factor can be approximated as
\begin{figure}
\centering
\includegraphics[scale=0.75]{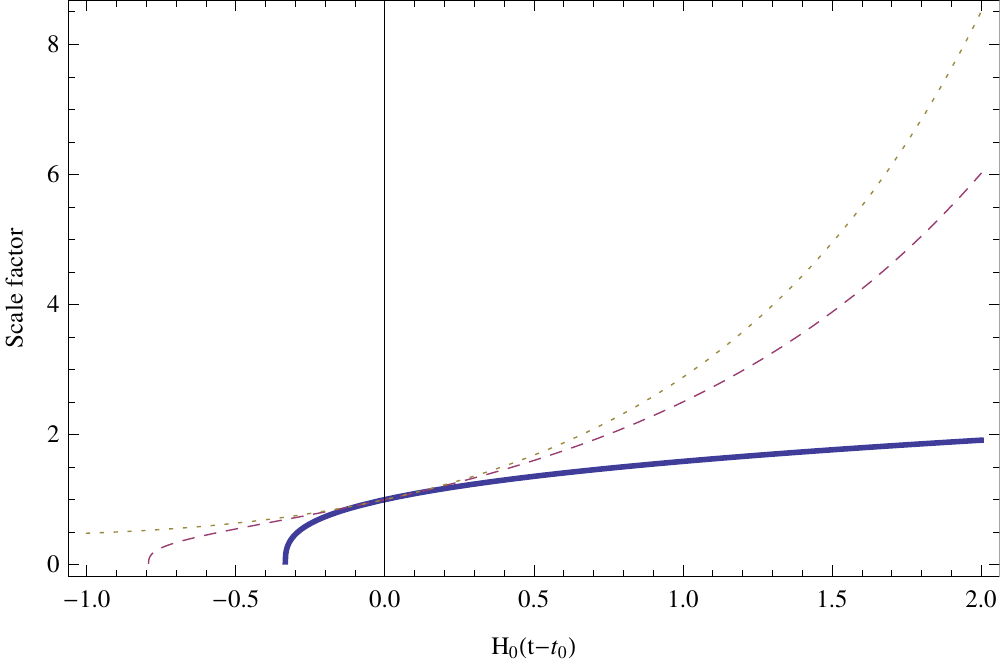}
\caption{Evolution of scale factor with time. Thick line corresponds
to $\bar\zeta =0.002$, dashed line corresponds to $\bar\zeta =5.25$
(best estimate) and dotted line  corresponds to $\bar\zeta=6.5.$}
\label{fig:avst}
\end{figure}
$a(t)\sim (1+3H_0(t-t_0))^{\frac{1}{3}},$ which implies decelerated
phase, while at large times the scale factor behaves as, $a(t)\sim
\exp(\frac{\bar\zeta}{6}H_0(t-t_0))$ showing that the universe follows
an accelerated epoch at a later time. The fig(\ref{fig:avst}) shows
evolution of scale factors at various choices of $\bar\zeta.$ From
the figure it is seen that the behavior of scale factor is different
for $\bar\zeta>6.$ If one finds the expression for the age of the
universe it is of the form
\begin{equation}
t_0-t_B=
\frac{2H_0^{-1}}{\bar\zeta}\log\left(\frac{6}{6-\bar\zeta}\right).
\end{equation}
For $\bar\zeta>6$ the age is not defined and consequently the
universe does not have a big bang. But for the cases $\bar\zeta<6$
the universe does have a big bang. For the best estimates of the
parameter the age of the universe is found to be around 10-12 Gy.

The equation of state of the bulk viscous fluid can be obtained
using the standard relation
\begin{equation}\label{eqn:EquationofState1}
\omega_z=-1-\frac{1}{3}\frac{d}{dx}(\ln {h^2})
\end{equation}
where $x=\ln a.$ Substituting for $h$ in
equation(\ref{eqn:EquationofState1}) using
equation(\ref{eqn:Hubblepara1}), we have
\begin{equation}\label{eqn:EquationofState2}
\omega_z=-1+\frac{2(6-\bar\zeta)}{\bar\zeta a^3+(6-\bar\zeta)}.
\end{equation}

\begin{figure}
\centering
\includegraphics[scale=0.75]{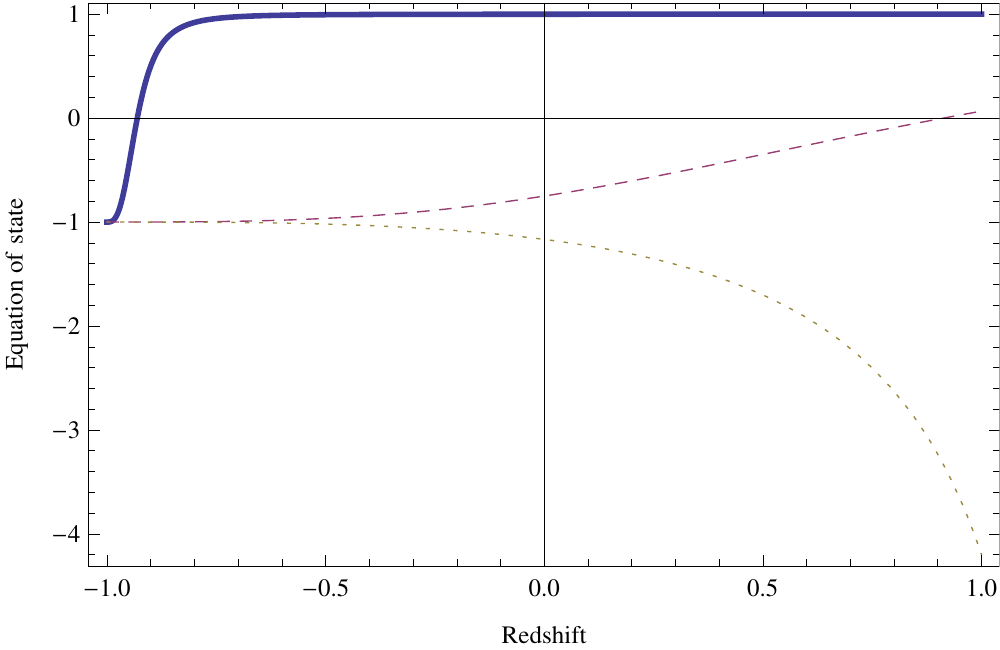}
\caption{Evolution of $\omega_z$ with respect to scale factor. Thick
line corresponds to $\bar\zeta=0.002$, dashed line corresponds to
$\bar\zeta=5.25$ and dotted line corresponds to $\bar\zeta=6.5$}
\label{fig:Omegavsz}
\end{figure}

The fig(\ref{fig:Omegavsz}) shows the variation of equation of state
parameter against the redshift at various choices of $\bar\zeta.$ In
the extreme future ($z \to -1$), $\omega_z \to -1$ and hence
corresponds to de Sitter universe.  Otherwise $\omega_z$ shows a
strong dependence on bulk viscous parameter. For small viscosity,
equation of state parameter remains $1$ but reduces to $-1$ in the
distant future. For $\bar\zeta<6$, like $\bar\zeta=5.25$, the best
estimated value, $\omega_z$ is positive for $z>1$. But in the
subsequent evolution, it reduces to negative values and finally
stabilizes at $-1$ as $z \to -1.$ This means that for $z<1$ the bulk
viscous Zel'dovich fluid mimics the quintessence nature for
$\bar\zeta>6,$ for instance when $\bar\zeta=6.5$ as in the figure,
when $\omega_z\leq -1$ always, which corresponds to a phantom
nature.

\begin{figure}
\centering
\includegraphics[scale=0.75]{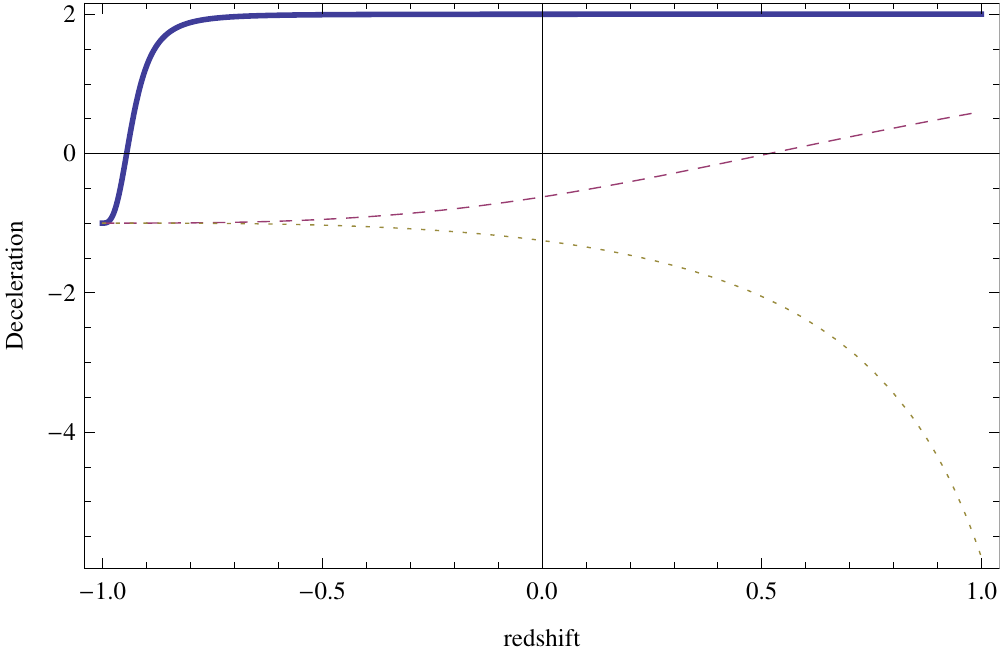}
\caption{Evolution of deceleration parameter with respect to red
shift. Thick line corresponds to $\bar\zeta=0.002$, dashed line
corresponds to $\bar\zeta=5.25$ and dotted line corresponds to
$\bar\zeta=6.5$} \label{fig:Decelerationvsz}
\end{figure}

We have also evaluated the deceleration parameter. The basic
equation of the deceleration parameter is
\begin{equation}
q=-1-\frac{\dot H}{H^2}
\end{equation}
Substituting for $H=H_0h$ from equation(\ref{eqn:Hubblepara1}) we
have
\begin{equation}
q=-1+\frac{3(6-\bar\zeta)}{\bar\zeta a^3+(6-\bar\zeta)}
\end{equation}
As seen in the figure(\ref{fig:Decelerationvsz}), the deceleration
parameter $q$ remains $-1$ for all possible $\bar\zeta$ in the
distant future. For a small value of $\bar\zeta$ the deceleration
remains at $2$ until a distant future when it drops down to $-1.$
For the best estimated value of $\bar\zeta=5.25,$ the switchover
from deceleration to acceleration takes place at about $z=0.52$ which
is closely in agreement with the observational constraints. The same
will proceed with ever increasing acceleration to an asymptotic
value $-1$ at a distant future. For values of $\bar\zeta>6.0,$ such
as the one indicated for $\bar\zeta=6.5$ as in the
figure(\ref{fig:Decelerationvsz}), q is always negative and is
increasing as the universe expands, saturating to -1 at $z \to -1$
in the future. So the final state of the universe in this model is a
de Sitter universe for any positive value $\bar\zeta.$
\section{Phase space perspective}
A convenient method to understand the global picture of the model is
to investigate into the equivalent phase space. For this, first, one
has to identify the phase space variables and be able to write down
the cosmological equations as a system of autonomous differential
equations. The critical points of these autonomous differential
equations can then be correlated to the cosmological solutions. If
the critical points were a global attractor, then the trajectories
of the autonomous system constructed near a critical point will
always be attracted towards it independent of the initial
conditions.

\subsection{Analysis of Zel'dovich fluid in two dimensional phase space
situation}\label{sssec:2DPhaseSpace}
 In this section the behavior of the system in the two
dimensional phase space with $h$ and $\Omega_z$ as the coordinates
is examined. The coupled differential equations
are
\begin{equation}\label{eqn:2DEquationsOFmotion1}
\dot h=H_0\left(\frac{\bar\zeta}{2}-3h\Omega_z\right)h=P(h,\Omega_z)
\end{equation}
\begin{equation}\label{eqn:2DEquationsOFmotion2}
\dot\Omega_z=-H_0\left[\left(6\left(\Omega_z-1\right)h-\bar\zeta\right)\Omega_z-\bar\zeta\right]=Q(h,\Omega_z)
\end{equation}
where $h=H/H_0$ and $\Omega_z=\rho_z/3H^2.$ By setting $\dot h=0$
and $\dot \Omega_z=0$, we obtain the following three critical points
 \begin{equation}\label{eqn:roots2D1}
(h_c,\Omega_{zc})=(0,1),(\frac{0.87667}{\Omega_z},\Omega_z),(0.87667,1).
\end{equation}

The first root corresponds to a static
universe, while the second root depends on the instantaneous value
of $\Omega_z$ and hence it is not a fixed point. The third root
$(h_c,\Omega_{zc})=(0.87667,1)$ corresponds to an expanding
universe dominated by Zel'dovich fluid.
 If the system is stable in the neighborhood of a critical
point, the linear perturbation in its neighborhood in phase space
decays with time. The perturbations  around the critical points must
satisfy,
\begin{equation}
\label{eqn:2Djacobian1}
\begin{bmatrix}
\dot\epsilon \\\dot \eta
\end{bmatrix}
=\begin{bmatrix}\left(\frac{\partial P}{\partial h}\right)_0
&\left(\frac{\partial P}{\partial \Omega_z}\right)_0
\\
\left(\frac{\partial Q}{\partial h}\right)_0 &\left(\frac{\partial
Q}{\partial \Omega_z}\right)_0
\end{bmatrix}
\begin{bmatrix}
\epsilon\\\eta
\end{bmatrix}
\end{equation}
Here $\epsilon$ and $\eta$ are perturbations in $h$ and $\Omega_z$
respectively in the neighborhood about a given critical point. The
corresponding Jacobian is
\begin{equation}\label{2Djacobian2}
\begin{bmatrix}\left(\frac{\partial P}{\partial h}\right)_0
&\left(\frac{\partial P}{\partial \Omega_z}\right)_0
\\
\left(\frac{\partial Q}{\partial h}\right)_0 &\left(\frac{\partial
Q}{\partial \Omega_z}\right)_0
\end{bmatrix}=H_0\begin{bmatrix}\left(\frac{\bar\zeta}{2}-6h\Omega_z\right)\ &-3h^2 \\ 6\Omega_z(1-\Omega_z)\ &\frac{\bar\zeta}{2}+3h(1-2\Omega_z) 
\end{bmatrix}\end{equation}
where the suffix $'0'$ implies the value at the critical point. The
secular equation leads to the eigenvalues describing the behavior of
the phase space trajectories near the equilibrium points.

The eigenvalues corresponding to the first critical point are found
to be $-368.2$ and $184.2$. As they are of opposite signs the
\begin{figure}[h]
\centering
\includegraphics[scale=0.75]{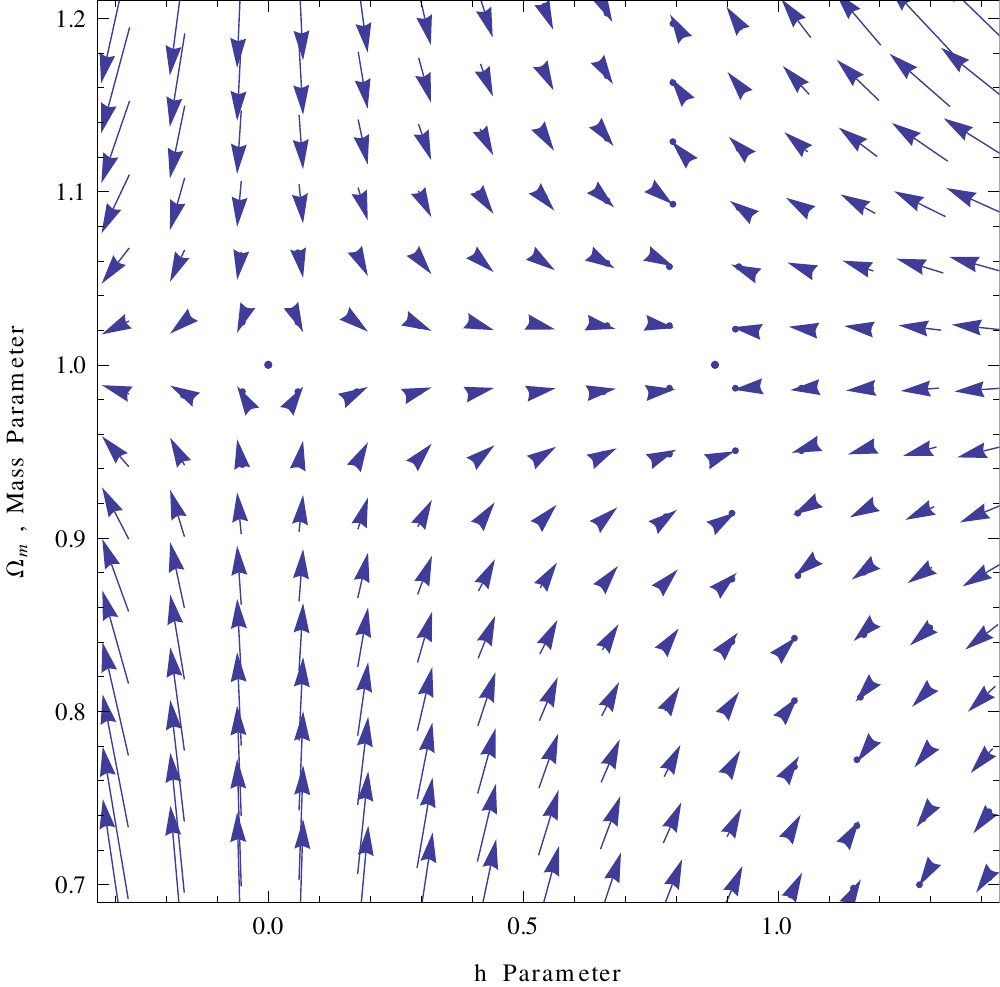}
\caption{Phase space structure around the critical points. The first critical point 
(0,1) is a saddle point. The
vector diagram clearly indicate that the trajectories approaching this point are 
repelled away and are finally converging on to the attractor critical point on the right.}
\label{fig:2Dcritpoint1}
\end{figure}
critical point is a saddle point and hence unstable. Depending on the initial conditions 
the nearby trajectories around this point may approach the saddle point, but repelled by it and finally 
approaching a possible stable attractor in the future.  As in the
figure \ref{fig:2Dcritpoint1} the trajectories are turning away from
the equilibrium point as and when they approach it and finally converges on the critical point shown on the right side of the plot.

The second critical point is not an isolated point, but varies with
$\Omega_z.$ As per the relationship between $h$ and $\Omega_z,$ it
represents a rectangular hyperbola with the axes $h=0$ and
$\Omega_z=0$ as asymptotes. The eigenvalues are found to be
$(-184.1, \frac{-161.394}{\Omega_z^2}).$ Since both the eigenvalues
are negative and real, the neighboring trajectories will converge on
to the hyperbola and hence the 'critical point' is a stable one. The
hyperbola along which the $\Omega_z$ dependent critical point moves
has the coordinate axes $h=0$ and $\Omega_z=0$ as the asymptotes,
$h=-\Omega_z$ as the directrix and $(0.9363,0.9363)$ as the focus.

\begin{figure}[h]
\centering
\includegraphics[scale=0.75]{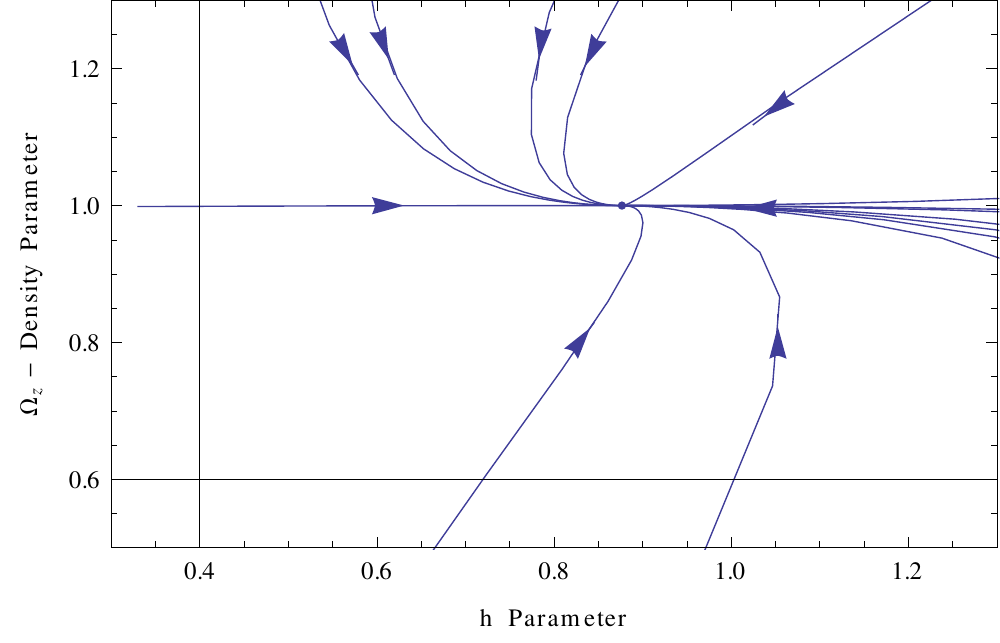}
\centering
\caption{Plot of phase space trajectories around the critical point
(0,87667,1). All trajectories are converging to the critical point
and hence it is stable.} \label{fig:2Dcont}
\end{figure}

The third critical point is $(h_c,\Omega_{zc})=(0.87667,1).$ It is
observed as in figure \ref{fig:2Dcritpoint1}, this critical point is a global
attractor. As the attempt was made to decouple the equation
(\ref{eqn:2DEquationsOFmotion1}) in the linear limit in the vicinity
of the critical point via equation (\ref{eqn:2Djacobian1}) it results
in the eigenvalues $-184.2$ and $0.0$.  The resulting two
eigenvalues clearly indicate the model is stable for all possible
initial conditions. It appears that
 the second eigenvalue $0$ is suggestive of absence of any isolated
critical point and rather a line segment as a continuous array of
critical points. However, a close examination of the vector field
\begin{figure}[h]
\centering
\includegraphics[scale=0.3]{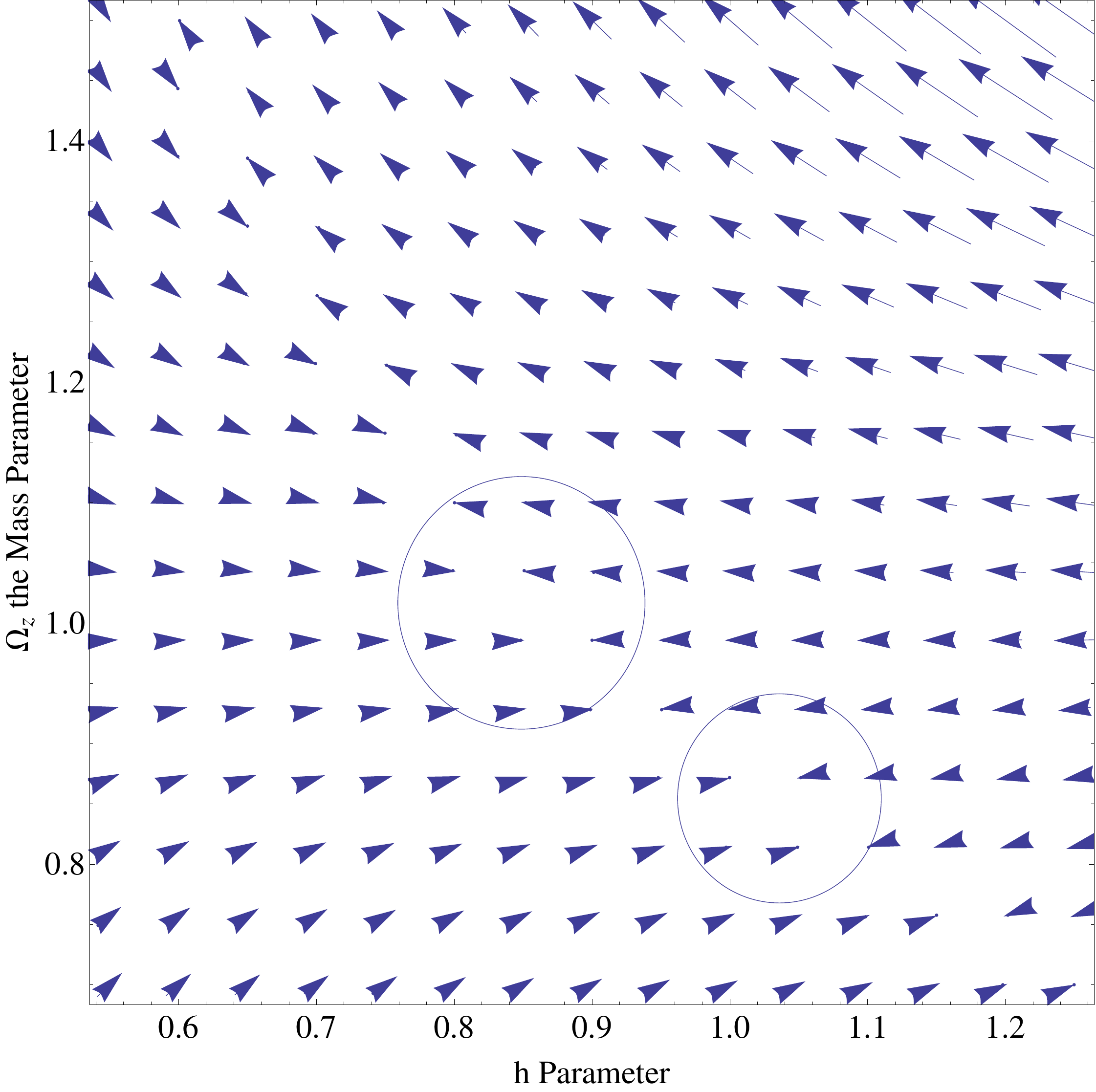}
\caption{Vector field plot of the phase space around the third
critical point (0,87667,1). The vectors encircled shows their
continuous tilt towards the critical point.}
\label{fig:2Dcritpoint4}
\end{figure}
plot as in figure \ref{fig:2Dcritpoint4} shows that the field
directions are invariably tilted, though slightly towards the
critical point as they approach what on low resolution seems to be a
straight line towards the isolated critical point. It is clearly
evident from the continuous plot phase space structure as shown in
figure \ref{fig:2Dcont}. So, in reality the isolated critical point
exists and the $'0'$ eigenvalue leads to a line segment as the best
fit close to the critical point. This is clear from the fact that
the straight line does not arise from the original procedure of
setting $\dot h=0$
and $\dot \Omega_z=0$ without the linear approximation and rather
results in an isolated point. So as we said earlier depending on the initial conditions the trajectories 
emanating from the surroundings of the saddle critical point are repelled away from it and they finally approache the stable critical point,
$(0.87667,1).$ This, by and large implyes the stability of the universe dominated by the bulk viscous Zel'dovich fluid.

\subsection{Analysis of Zel'dovich fluid in the three dimensional phase space situation}\label{sssec:3DPhaseSpace}
 In the case of the present model, the analysis is better
 realistic when we include, besides the Zel'dovich fluid, the conventional radiation
 also. The first Friedmann equation becomes
\begin{equation}\label{eqn:Friedmann3D}
 3H^2=\rho_z+\rho_\gamma
 \end{equation}
 where $\rho_\gamma$ is the radiation density. The conservation
 equation for the radiation component by assuming a pressure
 $p_\gamma=\rho_\gamma/3$, is
 \begin{equation}\label{eqn:conservationRad}
 \dot\rho_\gamma+4H\rho_\gamma=0
 \end{equation}.

The phase-space variables are $h$, $\Omega_z$ and $\Omega_\gamma$
among which the third parameter is $\Omega_\gamma=\rho_\gamma/3H^2.$
The dynamical equations for these parameters are represented by the
coupled differential equations
\begin{equation}\label{eqn:equationOFmotion1}
\dot h=P(h,\Omega_z,\Omega_\gamma)=\left((3\Omega_z+2\Omega_{\gamma})h-\frac{\bar\zeta}{2} \right)h
\end{equation}
\begin{equation}\label{eqn:equationOFmotion2}
\dot
\Omega_z=Q(h,\Omega_z,\Omega_\gamma)=\left[2\left((3\Omega_z+2\Omega_{\gamma})h-\frac{\bar\zeta
}{2}\right)-6h\right]\Omega_z+\bar\zeta
\end{equation}
and
\begin{equation}\label{eqn:equationOFmotion3}
\dot
\Omega_{\gamma}=R(h,\Omega_z,\Omega_\gamma)=\left[2\left((3\Omega_z+2\Omega_{\gamma})h-\frac{\bar\zeta
}{2}\right)-2h\right]\Omega_{\gamma}.
\end{equation}
The critical points are obtained by setting
\begin{equation}\label{eqn:criticalpointsset} \dot h=0,\\\
\dot\Omega_z=0,\\\ \dot\Omega_{\gamma}=0.
\end{equation}
and they are
\begin{equation}\label{eqn:criticalpoints2}
(h_c,\Omega_{zc},\Omega_{{\gamma}c})=(\frac{0.87667}{\Omega_{z}},  \Omega_{z},0)
; \, (0,1,0); \, (1,0.87667,0)
\end{equation}
out of which, the first mentioned is not fixed, having $h$ inversely
proportional to the instantaneous value of $\Omega_z$.  The second critical point corresponds to static universe and third 
one corresponds to an expanding universe dominated by bulk viscous Zel'dovich fluid. It is to be noted that there is no critical point 
corresponds to a radiation dominated phase. The
stability of the equilibrium points in the case of these three
 critical points are obtained (this time in the 3D phase-space case), once again by looking at the behavior of phase-space
trajectories close to them and generated due to different initial
conditions. The coupled differential equations in the linear limit
in matrix representation, in the neighborhood of the equilibrium
points are
\begin{equation}\label{eqn:Perturbation1}
\begin{bmatrix} \dot\epsilon \\ \dot \eta \\ \dot \nu
\end{bmatrix}
=\begin{bmatrix}\left(\frac{\partial P}{\partial h}\right)_0
&\left(\frac{\partial P}{\partial \Omega_z}\right)_0
&\left(\frac{\partial P}{\partial \Omega_\gamma}\right)_0\\
\left(\frac{\partial Q}{\partial h}\right)_0 &\left(\frac{\partial
Q}{\partial \Omega_z}\right)_0&\left(\frac{\partial Q}{\partial
\Omega_\gamma}\right)_0\\
\left(\frac{\partial R}{\partial h}\right)_0 &\left(\frac{\partial
R}{\partial \Omega_z}\right)_0 &\left(\frac{\partial R}{\partial
\Omega_\gamma}\right)_0
\end{bmatrix}
\begin{bmatrix}
\epsilon\\ \eta\\ \nu
\end{bmatrix}\end{equation}
where $\dot\epsilon$, $\dot\eta$ and $\dot\nu$ are first order
perturbation terms of $P(h,\Omega_z,\Omega_\gamma)=\dot h$,
$Q(h,\Omega_z,\Omega_\gamma)=\dot\Omega_z$ and
$R(h,\Omega_z,\Omega_\gamma)=\dot\Omega_{\gamma}$ respectively,
$\epsilon(t)$, $\eta(t)$ and $\nu(t)$ being the first first order
linear perturbation terms of $h,\Omega_z$ and $\Omega_\gamma$
respectively. The square matrix term in the equation
(\ref{eqn:Perturbation1}) is the Jacobian evaluated at the critical
point. We then decouple the differential equations
(\ref{eqn:Perturbation1}) by means of the secular equation and in
the process the eigenvalues corresponding to the equilibrium point
$(\frac{0.87667}{\Omega_{zc}},\Omega_{zc},0)$ are obtained as
\begin{equation}
\lambda_1=\frac{0.5}{\Omega_z}\left(-368.2+365.57\Omega_z-\sqrt{135571-
269206\Omega_z+13364\Omega_z^2}\right), \end{equation}
\begin{equation}
\lambda_2=\frac{0.5}{\Omega_z}\left(-368.2+365.57\Omega_z+\sqrt{135571-
269206\Omega_z+13364\Omega_z^2}\right) \end{equation} and
\begin{equation}
\lambda_3=\frac{245.466}{\Omega_z}, \end{equation} all depending on
the instantaneous value of $\Omega_z.$ The critical point in this case
drifts with the variation of $\Omega_z$ along a rectangular
\begin{figure}[ht]
\centering
\includegraphics[scale=0.75]{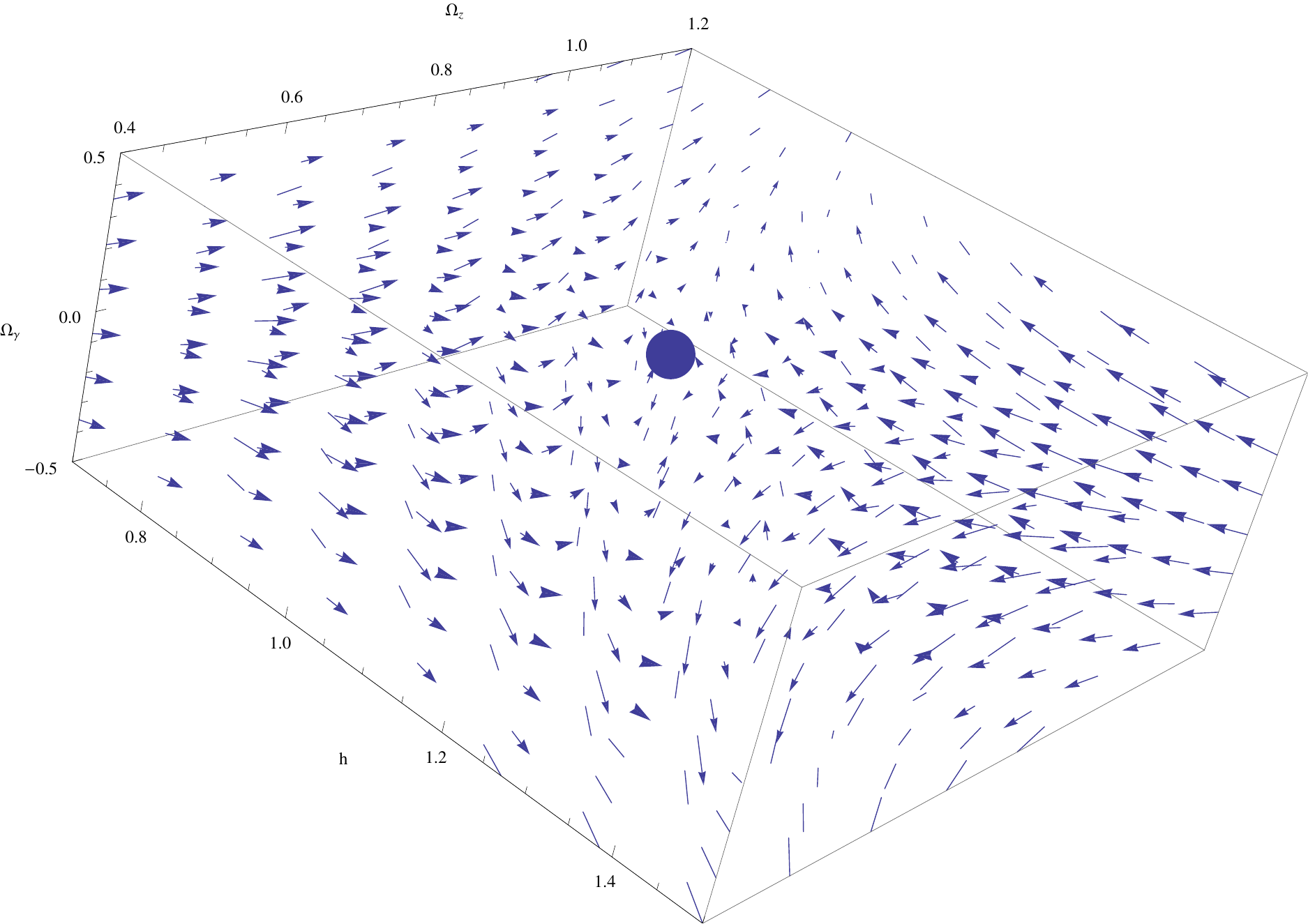}
\caption{Vector field plot of the phase-space structure around the critical point (1,0.87667,0)}
\label{fig:3Dcritpoint2}
\end{figure}
hyperbola on the $h-\Omega_z$ plane; with the details of the
hyperbola same as in the case of the second critical point in the
section \ref{sssec:2DPhaseSpace}. The eigenvalues indicate the phase
space trajectories corresponding to various initial conditions move
away from the critical point and hence no stable situation. Even the
case where $\Omega_z=0$ the eigenvalues are such that, $\lambda_1$ is negative,
$\lambda_2=0$ and $\lambda_3$ positive, and so stable solution 
solution is implied.

The second critical point $(0,1,0)$ has the eigenvalues $-368.2,$
$368.2$ and $184.1$ and the third critical point $(1,0.87667,0)$ has
the same to be $-403.778, 280.0$ and $167.878.$ There is one
negative eigenvalue and there are two positive eigenvalues for each critical
points which again means there is no stability for the equilibrium
points. This means the phase space trajectories are not attracted by
any of the critical points in the three dimensional case. For example the vector field plot 
as in the figure \ref{fig:3Dcritpoint2} clearly indicates how the phase-space trajectories corresponding to 
various initial conditions are repelled away, rather than being attracted to the second critical point.
 So
none of the critical points in this case corresponds to a radiation dominated phase and even the existing critical points are not stable also.
In fact the third critical point which corresponds to a Zel'dovich fluid dominated one, since it is unstable, it can be concluded that the inclusion of the 
radiation component may lead to a complete break down of the model. The bulk viscous coefficient is taken as a constant in the present study. Since it is 
a transport coefficient it may depend on the velocity of the fluid component also. Such a velocity dependent bulk viscous coefficient may be checked for consistency 
of a prior radiation dominated phase and that we reserve for a further work.

\section{Conclusion}
In this paper we have considered a flat universe consisting of bulk
viscous Zel'dovich fluid. The viscosity parameter was incorporated
as per the Eckart's formalism. We have evaluated the evolution of
the Hubble parameter. The model was constrained with SNe Ia data to
evaluate the bulk viscous coefficient $\bar\zeta=5.25\pm0.14$ the
present value of Hubble parameter $H_0=70.20\pm 0.58.$ The behavior of the
resulting scale factor shows that the model predicts a late
acceleration in the expansion of the universe. Hence the bulk
viscous Zel'dovich fluid can mimic the role of the conventional dark
energy.

We also studied the model to analyze the stability of the solutions
corresponding to various scenarios using the phase space analysis
method. We first analyzed the two dimensional phase-space behavior, where
the contribution due to radiation is neglected and
found that there is a past unstable saddle critical point corresponding to a static universe. The phase-space trajectories originating 
form the vicinity of this saddle like point are repelled away from it and are proceeded towards the stable critical point corresponding to 
an expanding universe dominated by Zel'dovich fluid. 

In the second instance we considered a three dimensional phase-space
case by incorporating the radiation component too. In this case no 
critical points are found corresponding to a prior radiation dominated phase and more over none of the existing critical points are stable.
  Hence the present model of the universe with bulk viscous Zel'dovich, in which bulk viscosity is characterized by a constant coefficient, first of 
all failed to predict a prior radiation dominated phase and secondly the very inclusion of the radiation makes the very model unstable.\\[0.3in]
\noindent\textbf{Acknowledgements} \\
We wish to thank IUCAA, Pune for the local hospitality during our
visits, where part of the work has been carried out. We are also
thankful to Prof. M Sabir and Prof. Varun Sahni for the discussions.



\begin{thebibliography}{99}
\bibitem{Perl1} Perlmutter S et al. 1999 {\it Astrophys. J.} {\bf 517} 565.
\bibitem{Reiss1} Reiss A G et al. 2004 {\it Astrophys. J.} {\bf 607} 665.
\bibitem{Hicken1} Hicken M et al. 2009 {\it Astrophys. J.} {\bf 700} 1097.
\bibitem{Shariff1}Shariff M and Abdul Jawad 2012 {\it Eur.Phys.J.C} {\bf 72} 2097.
\bibitem{Koivisto1}Koivisto T and Mota D F 2006 {\it Phys. Rev. D} {\bf 73} 083502.
\bibitem{Daniel1}Daniel S F 2008 {\it Phys. Rev. D} {\bf 77} 103513.
\bibitem{Fedeli1}Fedeli C, Moscardini L and Bartelmann M 2009 {\it Astron.Astrophys.} {\bf 500} 667.
\bibitem{Komatsu1}Komatsu E et al. 2011 {\it Astrophys. J. Suppl.}{\bf 192} 18.
\bibitem{Larson1}Larson D et al. 2011 {\it Astrophys. J. Suppl.}{\bf 192} 16.
\bibitem{Refregier1} Refregier A 2003 {\it Ann. Rev. Astron. Astrophys.}{\bf 41} 645.
\bibitem{Tyson1} Tyson J A, Kochanski G P, Del Antonio I P 1998 {\it Astrophys.J.}{\bf 498} L 107.
\bibitem{Allen1} Allen S W, Fabian A C, Schmidt R W, Ebeling H 2003 {\it Mon.not.R.Astron.Soc.}{bf\ 342} 287.
\bibitem{Zwicky1}Zwicky F 1933 {\it Hele. Phys. Acta} {\bf 6} 110.
\bibitem{Rubin1}Rubin V C, Ford W K J 1970 {Astrophys. J.}{bf\ 159} 379.
\bibitem{Dutta1}Dutta S, Hsu S D H, Reeb D, Sherrer R J 2009 {\it Phys.
Rev D} {\bf 79} 103504.
\bibitem{Zel'dovich1}Zel'dovich Ya B 1962 {\it Sov. Phys. JETP} {\bf 14}
11437.
\bibitem{Steile1} Steili R, Boeckel T, Schaffner-Bielich J 2010 {
\it Phys. Rev. D} {\bf 81} 123513.
\bibitem{Horava1} Horava P 2009 {\it Phys. Rev. D} {\bf 79} 084008.
\bibitem{Kiritsis1} Kiritsis E, Kofinas G 2009 {\it Nucl. Phys. B}
{\bf 821} 467.
\bibitem{Sotiriou1} Sotiriou T P, Visser M, Wuinfurtner S 2009 {\it
JHEP}{\bf 0910} 033.
\bibitem{Bogdanos1} Bogadanos C, Saritakis E N 2010{\it Class.
Quant. grav.}{\bf 27} 075005.
\bibitem{Barrow1} Barrow J D 1986 {\it Phys. Lett. B} {\bf 180} 335.
\bibitem{Fernandes1} Fernandez-Jambrina L, Gonzalez-Romero L M 2002{\it Phys. Rev. D} {\bf 66} 024027.
\bibitem{Fernandes2} Fernandez-Jambrina L 1997{\it Class. Quant. Grav.} {\bf 14} 3407.
\bibitem{Dutta2} Dutta S, Scherrer R J 2010{\it Phys. Rev.  D} {\bf 82} 083501.
\bibitem{Wilson1} Wilson J R, Mathews G J and Fuller 2007{\it Phys. Rev.  D} {\bf 75} 043521.
\bibitem{Ilg1} Ilg P and Ottinger H C 2000{\it Phys. Rev.  D} {\bf 61} 023510.
\bibitem{Zimdahl1} Zimdahl W 1996{\it Phys. Rev.  D} {\bf 53} 5483.
\bibitem{Paddy1} Padmanabhan T and Chitale S 1987{\it Phys. Lett. A} {\bf 120} 433.
\bibitem{Avelino1} Avelino A and Nucamendi U 2010 {\it JCAP} {\bf 08} 009.
\bibitem{Avelino2} Avelino A, Garcia-Salcedo R, Gonzalez T ,
Nucamendi U and Quirose I 2013 {\it JCAP} {\bf 08} 12
\bibitem{Athira1} Athira S and Titus K Mathew 2015 {\it Eur. Phys. J. C} {\bf 75} 348
\bibitem{Titus1} Titus K Mathew, Aswathi M. B. 2014{\it Eur. Phys. J. C} {\bf 74} 3188.
\bibitem{Masso1} Masso E, Rota R 2003{\it Phys. Rev. D} {\bf 68} 123504.
\bibitem{Eckart1} Eckart C 1940{\it Phys. Rev} {\bf 58} 919.
\bibitem{Landau1} Landau L D and Lifshitz E M {\it Fliuid Mechanics}(Adsiison-Wesley, 1958).
\bibitem{Hiscock1} Hiscock W A, Lindblom 1985 {\it Phys. Rev. D} {\bf 31} 825.
\bibitem{Israel1} Israel W 1976 {\it Ann. Phys.}(NY){\bf 100} 310.
\bibitem{Israel2} Israel W and Stewart J M 1979 {\it Ann. Phys.}(NY){\bf 118} 341.
\bibitem{Israel3} Israel W and Stewart J M 1979 {\it Proc. R. Soc. Lond. A}{\bf 365} 43.
\bibitem{Kremer1} Kremer G M and Devecchi F P 2003 {\it Phys.Rev. D}{\bf 67} 047301.
\bibitem{Hu1} Hu M G and Meng X H 2006 {\it Phys.Lett. D}{\bf 635} 186.
\bibitem{Ren1} Ren J and Meng X H 2006 {\it Phys.Lett. D}{\bf 633} 1.
\bibitem{Hiscock2} Hiscock W A and Salmonson J 1991 {\it Phys.Rev. D}{\bf 43} 3249.
\bibitem{Kovalsky1} Kovalsky M et al. 2008 {\it Astrophys. J.}{\bf 686} 749.



\end{thebibliography}
\end{document}